\documentclass[aip,
 amsmath,amssymb,
 reprint,
]{revtex4-1}

\usepackage{graphicx}
\usepackage{dcolumn}
\usepackage{bm}
\usepackage[utf8]{inputenc}
\usepackage[T1]{fontenc}
\usepackage{etoolbox}
\usepackage{hyperref}
\usepackage{url}

\makeatletter
\def\@email#1#2{%
 \endgroup
 \patchcmd{\titleblock@produce}
  {\frontmatter@RRAPformat}
  {\frontmatter@RRAPformat{\produce@RRAP{*#1\href{mailto:#2}{#2}}}\frontmatter@RRAPformat}
  {}{}
}%
\makeatother
\begin{document}

\preprint{AIP/123-QED}

\title[Chemellia: Atomistic SciML]{Chemellia:\\An Ecosystem for Atomistic Scientific Machine Learning}
\author{Anant Thazhemadam}
\author{Dhairya Gandhi}%
 \affiliation{Julia Computing}

\author{Venkatasubramanian Viswanathan}
\author{Rachel C. Kurchin}
\affiliation{Carnegie Mellon University}
\email{rkurchin@cmu.edu}

\date{\today}

\begin{abstract}
Chemellia is an open-source framework for atomistic machine learning in the Julia programming language. The framework takes advantage of Julia's high speed as well as the ability to share and reuse code and interfaces through the paradigm of multiple dispatch. Chemellia is designed to make use of  existing interfaces and avoid ``reinventing the wheel'' wherever possible. A key aspect of the Chemellia ecosystem is the ChemistryFeaturization interface for defining and encoding features -- it is designed to maximize interoperability between featurization schemes and elements thereof, to maintain provenance of encoded features, and to ensure easy decodability and reconfigurability to enable feature engineering experiments. This embodies the overall design principles of the Chemellia ecosystem: separation of concerns, interoperability, and transparency. We illustrate these principles by discussing the implementation of crystal graph convolutional neural networks for material property prediction.
\end{abstract}

\maketitle

\section{Introduction}

With the increase in available computational power as well as the democratization of machine learning (ML) methods, application of ML to atomistic problems is rapidly gaining popularity~\cite{butler2018machine,prezhdo2020advancing,morgan2020opportunities,bedolla2020machine,margraf2023science}. Typically, the ML is interfaced with some simulation technique, such as density functional theory (DFT), molecular dynamics (MD), etc. This interfacing may be in the form of: (i) \textbf{Surrogatization}, replacing simulation entirely in order to run drastically faster, usually in cases where it is prohibitively expensive to run the full model the number of times that would be necessary, such as in design/discovery of new materials including electrocatalysts~\cite{Zhong2020}, high-entropy alloys~\cite{Wen2019}, intermetallics~\cite{Zhang2020}, and 2D topological insulators~\cite{Schleder2021}, among many others; (ii) \textbf{Acceleration}, where some expensive step is accelerated with the help of ML but the final result is still physically validated, such as generating ansatze for for quantum Monte Carlo~\cite{Pfau2020} or guiding DFT-based structural optimization~\cite{musielewicz2022finetuna}; or (iii) \textbf{Augmentation}, in which a core aspect of the simulation is replaced with an ML model to improve its performance/accuracy, such as ML potentials in MD~\cite{behler2007generalized} or ML functionals in DFT~\cite{Kalita2021, Kirkpatrick2021}.

In any of these applications, it is important to be able to interact with physically meaningful representations of the data that are ingestible by the ML model and annotated with relevant features. The representation of the data (usually a crystal or molecular structure) can be fully three-dimensional, or some simplified representation such as a graph. This representation is then \textit{featurized} before being fed into the model. Featurization may be as simple as annotation with an atomic or isotopic identity, or may explicitly impute other information about atoms, bonding environments, or overall structural ``fingerprints.'' There is a large variety of featurization and ML modeling approaches that have been utilized, but these are often implemented in a ``one-off" way, without much concern for interoperability or extensibility. These practices can also contribute to growing concerns about reproducibility in applications of ML to STEM problems~\cite{kapoor2022leakage}.

There are several existing efforts to create broader frameworks, each of which has its own focus area and functionality goals. For example, DeepChem~\cite{Ramsundar-et-al-2019} is a widely-used package aimed primarily at molecular systems (e.g. drug discovery) that supports a wide variety of featurizations and models through a common interface. Automatminer~\cite{dunn2020benchmarking}, in contrast, is substantially more materials-focused and aims to be very ``plug-and-play,'' allowing the user to abstract away specifics of model architecture, hyperparameters, and training procedure. Meanwhile, SchNetPack~\cite{Sch_tt_2018, https://doi.org/10.48550/arxiv.2212.05517} focuses specifically on incorporating newer developments in deep neural architectures that incorporate physical symmetries, e.g. via equivariant message-passing, and also incorporates GPU-accelerated molecular dynamics directly. As a final example, DScribe~\cite{himanen2020dscribe} provides an interface to many common featurization schemes, such as Coulomb matrices~\cite{rupp2012fast}, atom-centered symmetry functions~\cite{behler2011atom}, or smooth overlap of atomic positions~\cite{bartok2013representing}.

It is well-established that the solution to a proliferation of standard approaches is usually not another standard approach.~\cite{munroe_2011} Therefore, we introduce a new framework here not with the goal that it will become universally adopted, but rather in the hopes that it might become the primary such framework \emph{in the Julia language}.
Separately, we also believe that Julia will come to represent an increasing share of scientific computing use cases. As is discussed in further detail below, the design of the language facilitates high performance as well as broad \emph{interoperability}, which is particularly crucial in the second two ML applications described above (acceleration and augmentation), wherein the ML model needs to interface directly with simulation.

The remainder of this document serves to justify why Julia is a promising language for atomistic ML, and also to introduce the principles of the Chemellia framework, in which we try to adopt established best practices as well as learn from prior design choices that haven't served the community as well. We finish with a ``showcase'' demonstrating a concrete implementation of the framework in the form of crystal graph convolutional neural networks and an associated featurization.

\section{Design Principles}
\subsection{Why Julia?}

In designing any computational framework, choice of programming language suffuses virtually every other design and implementation decision~\cite{spinellis2006choosing}. Like human languages, some programming languages can express certain concepts or structures in more concise/intuitive ways than others. Unlike human languages, programming languages can also offer orders of magnitude of separation in performance, lending potentially tremendous weight to this decision~\cite{bench}.

The Julia language is a relatively young (decade-old) entrant to the scientific computing scene, but one with potential to be a major part of its future. This is largely because it achieves a kind of Pareto optimality between expressiveness and performance, solving the so-called ``two-language problem,'' i.e. the common paradigm of prototyping in a high-level but low-performance language such as Python and then re-implementing in a lower-level fast language such as C~\cite{perkel2019julia}.

The primary paradigm of Julia is multiple dispatch, which has emerged as an effective solution to the age-old expression problem \cite{wadler_1998}, allowing both users and developers to extend interfaces and types nearly orthogonally. 
For example, the JuliaGraphs~\cite{Graphs2021} ecosystem defines an interface for custom graph types; by dispatching a relatively minimal set of functions (for example, to return the number of nodes in a graph or whether there exists an edge between two given nodes), many sophisticated functionalities in the ecosystem's packages (e.g. centrality measures, algorithms for graph traversal, etc.) will ``just work'' on a user-defined graph type.


Julia also has strong, language-wide support for automatic differentiation (AD) in both forward and reverse mode via a variety of packages \cite{DBLP:journals/corr/abs-2109-12449}. It allows both end users and libraries alike to specify evaluation rules for their custom types via multiple dispatch through the common interface provided by ChainRules.jl \cite{frames_white_2023_7669643}. This eases the burden on one library internally supporting AD for all of its types. This is in sharp contrast to AD frameworks such as JAX~\cite{jax2018github}, which need to re-implement differentiable versions of any functions they wish to support (cf. \verb|jax.numpy|).

\subsection{Separation of Concerns}
Separation of concerns is a fundamental software design principle~\cite{mili2004understanding}. Broadly, it relates to \emph{modularity} of software, i.e., a given section of code should only deal with the information relevant for it to do its intended task. This principle fits naturally within the multiple dispatch paradigm of Julia, and is especially important in creating functional, composable interfaces.

Within Chemellia, separation of concerns emerges at a few levels.
First, as is common practice in the Julia ecosystem broadly, we generally aim for smaller and more scope-focused modular packages that use generic interfaces. 
Second, separation of concerns also helps in defining the logical constituents of a framework, as well as promoting more maintainable and extensible code.
In Chemellia, this manifests primarily in the logical boundary between method-level feature-engineering functions (e.g. encoding/decoding operations) and the actual model architecture composition itself (which can sometimes include featurization operations as part of its action as well). 



To elucidate these ideas in a more concrete way, we can consider a few specific model architectures.
Consider CGCNN \cite{cgcnn}, the earliest work that  demonstrated the effectiveness of GCN's for crystalline structures, as a baseline. 
In the original CGCNN, neighbor relationships between atoms were represented by the presence or absence of an edge between the corresponding nodes (or multiple edges for periodic images). However, GeoCGCNN~\cite{Cheng2021} introduced manual features which included topological distance and spatial distance as well. The architectural composition of the model itself remains largely the same in both cases. However, it is the feature engineering operation that sets these two models apart.
In contrast, consider MT-CGCNN \cite{DBLP:journals/corr/abs-1811-05660}, which is essentially CGCNN coupled with multi-task learning. Here, the model architecture itself is modified, but the way the features are represented remains the same. 

Effectively separating out concerns in this way also allows us to realize how model architectures are generalized from other domains; that is, which layers or featurizations could (or should) be composable/interoperable.
For instance, different pooling mechanisms, such as EigenPooling \cite{DBLP:journals/corr/abs-1904-13107}, can be effectively applied to other GCN's \cite{PredictionOfMP}. Designing the framework with this outlook of separation of concerns also helps abstract away the model architectures from the intricacies and specifics associated with the domain itself, to a great extent, as well as maximize interoperability (i.e. minimize the need to re-implement the same functionality over and over again to fit with different architectures).

\subsection{Interoperability}

Julia, powered by the multiple dispatch paradigm, enables interoperability from a  fundamental language level.
The scope of interoperability Chemellia and its libraries offer, and how they build upon that offered by the language itself, can be explored along different dimensions, which we explore briefly in this section.

Implementing the right abstractions helps introduce interoperability at the abstract level. This ensures that logically distinct concepts supported by the framework, such as featurizers, different encoding mechanisms, types of layers, etc, have clear and well-defined boundaries. This also ensures that users have an easier time navigating the framework and using it.

\begin{figure}
    \centering
    \includegraphics[width=0.48\textwidth]{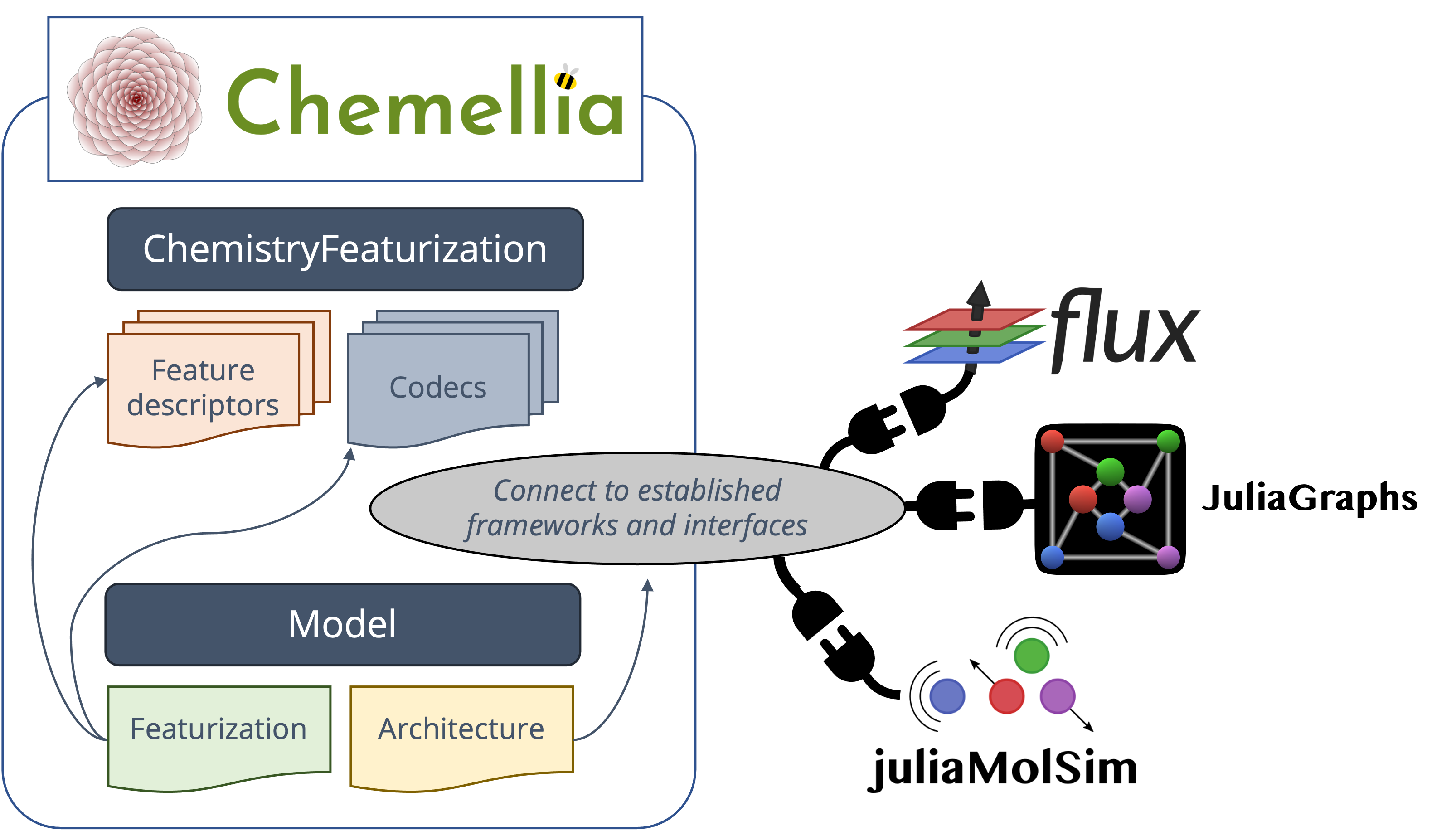}
    \caption{Overview of components of the Chemellia framework, and connections between them.}
    \label{fig:overview}
\end{figure}

Once these abstractions are well defined, the next task is implementation. While implementing a scheme of abstraction and guaranteeing interoperability between these abstract concepts might seem obvious and trivial, more often than not, the code begins to slowly evolve in a manner that blurs the lines between the abstractions and further complicates the relationships between them, especially if these abstractions are not clearly defined. This causes friction both while developing and using the framework. However, when carefully done, it empowers users with the freedom to mix and match between different components, in a plug-and-play fashion, within reasonable limits.
Furthermore, in cases where logical components that are canonically correct on their own are not immediately compatible with one another for practical reasons, ChemistryFeaturization (thanks to the structure of the Julia Language) typically makes it easy to write simple ``adapters.'' One example of this is the idea of a Codec, described in more detail in the Showcase below (see also Figure~\ref{fig:overview}). 

Specifics of implementation can also have a major role in interoperability at the ecosystem level. This relates to how well types and components from other packages across the ecosystem compose with the framework. In this context, the ideal user experience is being able to quickly integrate relevant components from appropriate packages and prototype, rather than spending time jumping through hoops trying to get the framework to integrate properly with said packages or vice versa. Because much of the Julia language is built around this kind of interoperability, it is generally very smooth to leverage other packages within the broader (open-source) Julia ecosystem. For example, Chemellia supports AtomsBase~\cite{Herbst_AtomsBase_jl_2023}, an interface designed to improve interoperability in specification of atomic geometries (for example, to pass between a simulation and ML model, or for visualization or file I/O). Another example, discussed in more detail below, is the ability to utilize model layers defined in GeometricFlux~\cite{GeometricFlux}.

In addition, Julia also provides excellent foreign function interface (FFI) support. This allows users to utilize more well-established ecosystems and frameworks for their specific requirements, if the need arises, through built-in methods like \verb|ccall| or packages like PythonCall~\cite{PythonCall.jl}, PyCall~\cite{PyCall.jl}, RCall, and JavaCall~\cite{JuliaInterop}. For instance, one of Chemellia's libraries, AtomGraphs, uses the \verb|pymatgen|~\cite{ong2013python} library via PythonCall for certain operations related to graph building.

\subsection{Transparency}
Transparency in code is valuable along myriad dimensions. It can also refer to a number of distinct concepts. For our purposes, we define transparency as the ease with which a user can understand what the code is doing. This is important to ensure that users know how to use the code in the first place, and also to allow them to validate correctness of results. 

The first part of this (how to use it) is largely enabled (along with thorough documentation, of course) by giving functions and types clear and descriptive names. When designing an interface such as ChemistryFeaturization that may have diverse concrete implementations, this can be particularly challenging, and the development team has in some cases had extensive discussions about the best name of a single entity in the codebase! 

The second part (validating correctness of results) comes primarily from readability of the source code itself. The vast majority of machine learning models ultimately boil down to matrix multiplication, along with a few other simple transformations (e.g. elementwise application of a nonlinearity). However, in many of the widely used implementations, it can be frustratingly difficult, given a model object, to ascertain which matrices are multiplied to produce the model's results! The Julia packages (in particular Flux~\cite{flux1,flux2}) upon which Chemellia is based largely avoid the massive type hierarchy often present in Python-based frameworks. Julia's metaprogramming tools are also important in code validation. For example, it is easy to use the \verb|@which| macro to find exactly what source code will be run in a given method invocation.

Another important aspect of transparency (that encompasses aspects of both of the prior two points) pertains specifically to featurization. Namely, how much information is actually encoded in a given featurization approach, and how much of the full input could be reconstructed from what has been encoded? The ChemistryFeaturization interface explicitly addresses this by exposing, along with the \verb|encode| function for computing a featurization of a given input structure, a \verb|decode| function that should invert that featurization procedure, to the extent possible.
(We note briefly here that in this context, we are using the words ``encode'' and ``decode'' specifically to mean translating from human-readable feature values to machine-readable representations and back again, respectively. This is a slightly narrower definition than, for example, the one implicit in the idea of an autoencoder.)
A common approach for encoding atomic features is a so-called ``one-hot'' scheme, where a single one in inserted in a string of zeros to represent the ``bin'' in which an encoded value falls. Of course, for categorical variables (e.g. group in the periodic table), this is a lossless encoding scheme. However, for continuous-valued ones (e.g. atomic radius), the choice of binning scheme determines the ``resolution'' with which this information is actually passed to the network. See Section~\ref{sec:featurization} and Figure~\ref{fig:featurization} below for more on this. 

In general, prioritizing transparency in all of the above senses helps code be only as complicated as it needs to be. By the same token, however, it is worth noting also that more transparency is not \emph{always} better, as it can become overwhelming and impede usability. Sometimes, the ability to control how much detail is abstracted away by an interface can be important, especially to users learning that interface for the first time. However, if the balance is struck properly, transparency should be an aid in the learning process. More transparent, less overwhelming designs also help enforce better separation of concerns, since now interfaces are built with the goal of exposing users who may be using a given functionality to only the relevant details, and thus compartmentalizing concepts more effectively. Overall, adopting these principles can help reduce the divide between users and developers, lowering barriers to usability, adoption, and (eventually) contribution.

\section{Showcase}
In order to demonstrate the principles we've outlined above, we will showcase a particular set of functionalities within Chemellia around building graph representations of crystals and models that can ingest them. This section closely mirrors a Pluto~\cite{pluto} notebook that can be found at \url{https://github.com/Chemellia/Chemellia_showcase} and run live on JuliaHub at \url{https://juliahub.com/ui/Notebooks/thazhemadam/Chemellia\%20Showcase/jcp_showcase.jl}, and the functionality can be found in the open-source packages ChemistryFeaturization, AtomGraphs, and AtomicGraphNets, all hosted (with documentation) on GitHub within the Chemellia organization and registered in the Julia Base registry.

\subsection{Graph Building}
The first step is to convert a standard representation of a crystal structure (e.g. a CIF or XYZ file) into a graph representation that can be used by the model we'll build later. We make use of the AtomsBase~\cite{Herbst_AtomsBase_jl_2023} interface to handle 3D crystal structure representations -- this abstract interface exposes a standard set of functions (e.g. \verb|position|, \verb|bounding_box|) for accessing information about these structures that may be stored in different types of data structures in different concrete implementations. It facilitates interoperability between packages for tasks like chemical simulation, file I/O, and visualization.

\begin{figure}
    \centering
    \includegraphics[width=0.4\textwidth]{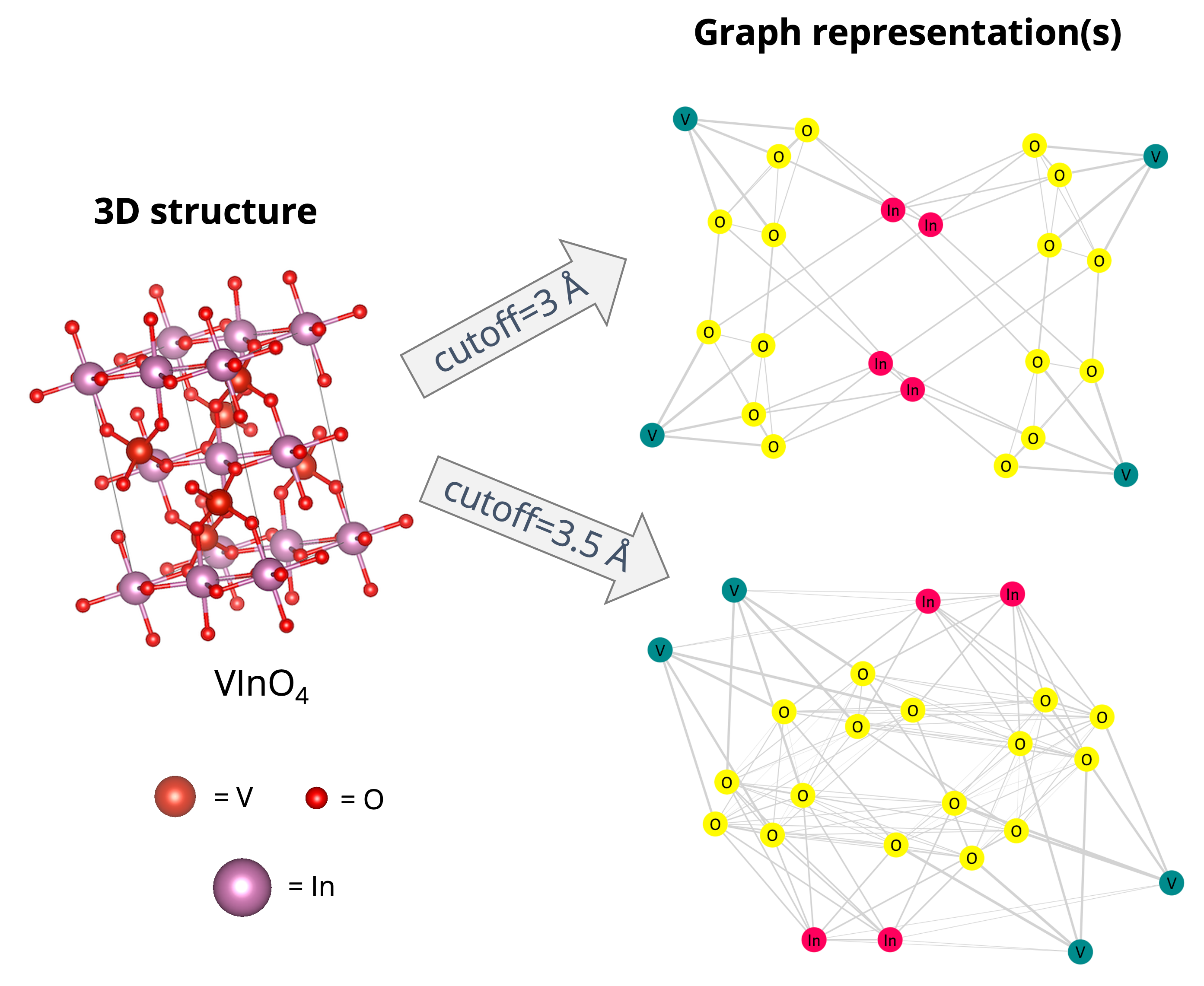}
    \caption{Demonstration of impact of neighbor list cutoff on connectivity of constructed graph: a smaller cutoff (above) leads to a sparser adjacency matrix (fewer graph edges).}
    \label{fig:graphviz}
\end{figure}

The \verb|AtomGraph| data type that we'll use for crystal graph representations is implemented in the AtomGraphs package within the Chemellia ecosystem. In contrast to some other atomic graph representations, an \verb|AtomGraph| is a weighted graph representation (in particular, it utilizes the \verb|SimpleWeightedGraph| type from the JuliaGraphs~\cite{Graphs2021} ecosystem/interface. 
The neighbor lists (i.e. other atoms/nodes to which to draw these weighted edges) can be chosen via setting distance/number cutoffs or (for periodic systems) by Voronoi tessellation, and the edge weights can be set by a user-defined function or one of several built-in options (such as exponential or inverse-square decay). In the notebook, the resulting graph topology from different choices of these options are explored in more detail. For example, Figure~\ref{fig:graphviz} shows how the neighbor cutoff affects connectivity of the graph. The graph visualizations were produced using  the GraphPlot package from JuliaGraphs.

\subsection{Featurization}
\label{sec:featurization}
With the structure representations (in this case graphs) constructed, they next need to be featurized -- that is, ``annotated'' with information about the constituent atoms, bonds, or groups thereof upon which the final material property prediction may depend (common choices include atomic number, valence, and electronegativity, among many others). The ChemistryFeaturization interface facilitates featurization and feature engineering through two primary abstractions:

\begin{itemize}
  \item Feature descriptors, which are used to uniquely describe a specific feature of an atom, bond, neighborhood, or entire structure. For this showcase, we will focus on \verb|ElementFeatureDescriptor|s, which describe an atom and require only its chemical identity to be defined.
  \item Codecs, which specify mechanisms for encoding and decoding data of a specific type. A common choice is so-called ``one-hot'' encoding, where a feature is represented as a bitstring with all zeros and a single one for the ``bin'' into which the value falls. In ChemistryFeaturization, this is accomplished via the \verb|OneHotOneCold| codec, which facilitates choice of binning scheme (e.g. linear vs. logarithmic scaling, number of bins), and allows easy encoding \emph{and decoding} so that a user can easily query the ``precision'' with which a continuous-valued feature is represented 
\end{itemize}

\begin{figure}
    \centering
    \includegraphics[width=0.45\textwidth]{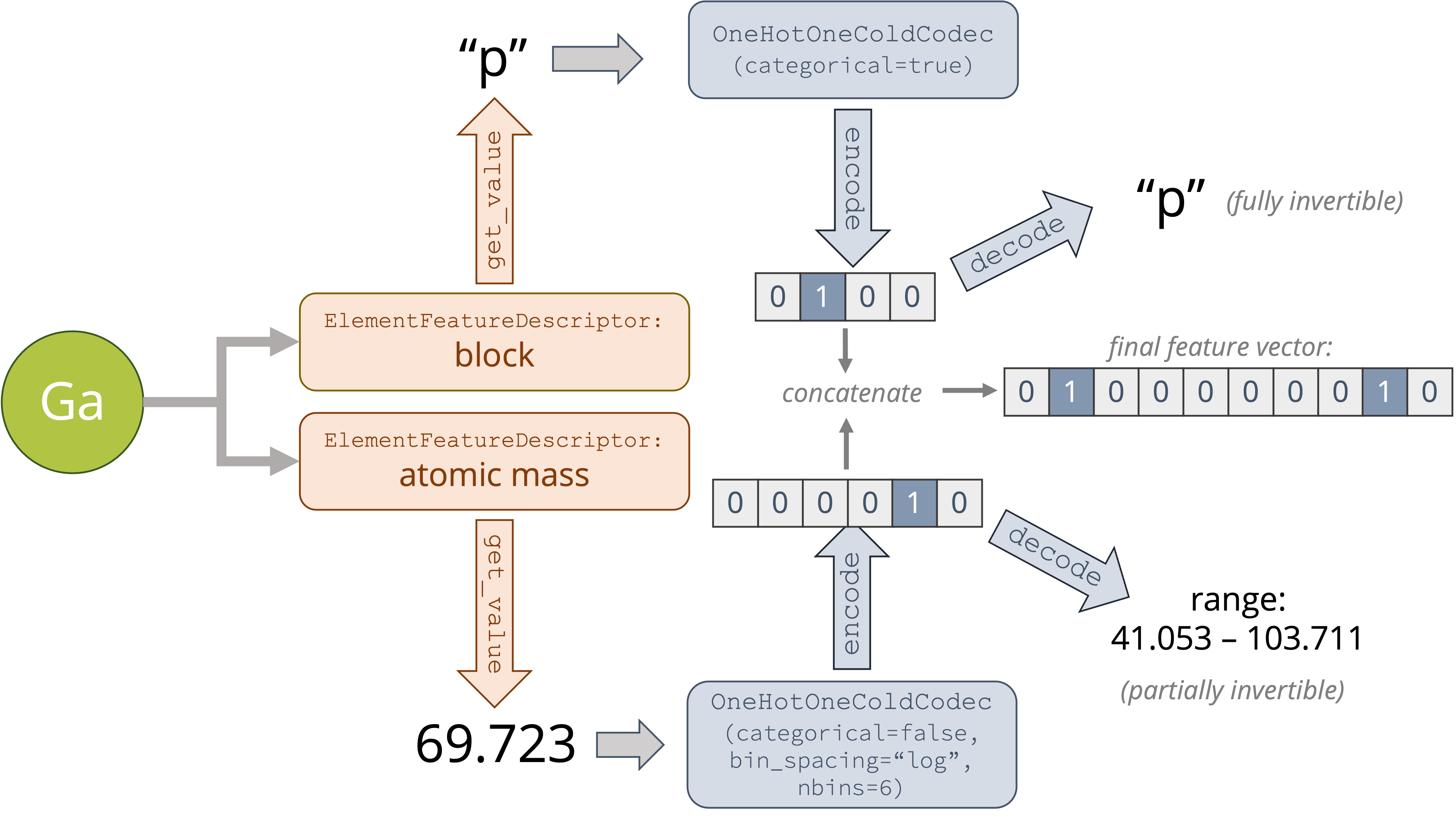}
    \caption{Schematic of encoding and decoding. The element gallium is fed as input to two different element feature descriptors: block (a categorical-valued feature), and atomic mass (a continuous-valued one). These are one-hot encoded and concatenated into a final node feature vector. Decoding is also demonstrated, showing how the ``resolution'' of encoding for continuous-valued features can easily be queried.}
    \label{fig:featurization}
\end{figure}

In this language, a ``featurization'' (which subtypes the \verb|AbstractFeaturization| type) comprises a set of feature descriptors and associated codecs, as well as a specification for how these encoded features should be combined in order to be ingested by a model. In \verb|GraphNodeFeaturization|, for example, we choose a set of \verb|ElementFeatureDescriptors| and associated \verb|OneHotOneCold| codecs, and the resulting node feature vectors are ``stacked'' into a feature matrix. In the demonstration notebook, this is illustrated with a featurization with just two features (one categorical and one continuous-valued): the block of the periodic table in which the element resides ($s$, $p$, $d$, or $f$) and the atomic mass of the element (See Figure~\ref{fig:featurization}). We also show the capacity to customize behavior in more detail as described above.

Since the featurization scheme, as a logical concept and a programming abstraction, is distinct from the actual data (e.g., the \verb|AtomGraphs|), we can, with minimal modification, apply the same featurization scheme to different values and types of representations as well. In addition, the separation of concerns between the feature descriptor and the codec, as well as the transparency enabled by enforcing the presence of a \verb|decode| function, facilitates feature engineering experiments by the ease of switching out or reconfiguring codecs and sets of feature descriptors independently.

The result of employing a featurization scheme onto a representation of the crystal is a \verb|FeaturizedAtoms| object, which encapsulates the featurization scheme, the original representation itself (which, within reasonable expectations, enables provenance and validates serialization), and the featurized representation obtained by encoding the featurization scheme onto the original representation. It is this \verb|FeaturizedAtoms| object that is fed into the model (see next section).

\subsection{Model Building}
AtomicGraphNets is a minimal implementation of a model conceptually similar to the original CGCNN~\cite{cgcnn}. However, as mentioned above, the graphs themselves have continuous-valued weights, and the convolution operation is accomplished via the graph Laplacian, allowing for fast performance in both the reverse (training) and forward directions.
AtomicGraphNets is built on the Flux.jl~\cite{flux1,flux2} ML framework, which in turn utilizes Zygote~\cite{Zygote.jl-2018} for AD support. In keeping with the discussion of interoperability above, the Flux stack has been widely used across a variety of applications that necessitate composability and interoperability between ``traditional'' ML model elements and other functionality -- for example, in state-of-the-art implementations of neural differential equations~\cite{diffeqflux}, wherein a neural network is coupled with ODE solvers for tasks such as model surrogatization or parameter estimation.


AtomicGraphNets defines some custom layers that can be easily composed with other Flux layers (or any differentiable function with an appropriate call signature). In the notebook, we demonstrate using these layers (via a convenience function) to build a model very similar to the standard CGCNN~\cite{cgcnn} architecture and show how to perform a training step.

Another package, GeometricFlux~\cite{GeometricFlux}, also provides implementations of a number of standard layers/operations common in graph-based ML generally. It is straightforward to dispatch the operations of these layers onto our \verb|FeaturizedAtoms| objects to take advantage of this functionality without needing to re-implement anything or copy code (this is shown in more detail in the notebook for two particular example layers: convolutional and pooling operations).


The functionality demonstrated here is just the beginning. Work is ongoing in implementing additional feature descriptors and codecs in ChemistryFeaturization, as well as new model architectures, and perhaps more importantly, supporting existing ones already implemented elsewhere.


\section{Conclusion}
Atomistic modeling in Julia (both data-driven and otherwise) is still in its early days. While a young language, Julia is nonetheless poised to become a major player in the future of scientific computing thanks to its speed and interoperability facilitated by multiple dispatch, but also due to the community of users and developers that has sprung up around it. We have built the framework for Chemellia to fit within these existing pieces (e.g. adopting interfaces such as AtomsBase) but also to be lightweight and adaptable so it can evolve along with the language and tooling. We believe that the emphasis on retaining provenance and ``decodability'' is is crucial for reproducible science and can also be of tremendous value to users learning these concepts for the first time -- making it easier to ``peek under the hood'' helps to encourage a new user to understand what is going on in the code rather than just how to use it.

To be sure, Chemellia ecosystem has functionalities that are yet to be ``fleshed out'' in detail, but we hope it can be a part of this bright future for atomistic modeling and, in particular, seamless interplay between data-driven approaches and traditional simulation. The Chemellia developer team is currently small, but open to growth as well as contributions of packages/functionality!

\begin{acknowledgments}
The information, data, or work presented herein was funded in part by the Advanced Research Projects Agency-Energy (ARPA-E), U.S. Department of Energy, under Award Number DE-AR0001211. The views and opinions of authors expressed herein do not necessarily state or reflect those of the United States Government or any agency thereof.

The authors also wish to acknowledge financial support from the Google Summer of Code Program, the Carnegie Mellon Manufacturing Futures Initiative Postdoctoral Fellowship, and the Molecular Sciences Software Institute Postdoctoral Fellowship.
\end{acknowledgments}

\bibliography{refs}

\end{document}